# COMPUTER ASSISTED PLANNING AND ORBITAL SURGERY: PATIENT-RELATED PREDICTION OF OSTEOTOMY SIZE IN PROPTOSIS REDUCTION

Vincent Luboz[1], Dominique Ambard[2], Pascal Swider[2], Frank Boutault[2,3], Yohan Payan[1]

1. TIMC Laboratory, UMR CNRS 5525, La Tronche, France
2. Biomechanics Laboratory EA 3697, Purpan University Hospital, Toulouse, France
3. Maxillofacial Surgery Department, Purpan University Hospital, Toulouse, France

Corresponding author:
Professor Pascal Swider
Biomechanics Laboratory EA 3697  (Director)
University Toulouse III - University Hospital Purpan
Place Dr. Baylac 31059 Toulouse cedex  - FRANCE
email: swider@cict.fr

**Abstract**

*Background.* Proptosis is characterized by a protrusion of the eyeball due to an increase of the orbital tissue volume. To recover a normal eyeball positioning, the most frequent surgical technique consists in the osteotomy of orbital walls combined with the manual loading on the eyeball. Only a rough clinical rule is currently available for the surgeons but it is useless for this technique. The first biomechanical model dealing with proptosis reduction, validated in one patient, has been previously proposed by the authors.

*Methods.* This paper proposes a rule improving the pre-operative planning of the osteotomy size in proptosis reduction. Patient-related poroelastic FE models combined with sensitivity studies were used to propose two clinical rules to improve the pre-operative planning of proptosis reduction. This poroelastic model was run on 12 patients. Sensitivity studies permitted to establish relationships between the osteotoemy size, the patient-related orbital volume, the decompressed tissue volume and the eyeball backward displacement.

*Findings.* The eyeball displacement and the osteotomy size were non-linearly related: an exponential rule has been proposed. The patient-related orbital volume showed a significant influence: a bi-quadratic analytical equation liking the osteotomy size, the orbital volume and the targeted eyeball protrusion has been established.

*Interpretation.* Two process rules derived from patient-related biomechanical FE models have been proposed for the proptosis reduction planning. The implementation of the process rules into a clinical setting is easy since only a sagittal radiography is required. The osteotomy size can be monitored using optical guided instruments.

# 1. Introduction

Proptosis is characterized by the increase of the volume of the orbital content mostly due to an endocrinal dysfunction (Saraux et al., 1987). The protrusion of the eyeball (Fig.1 (a)) induces aesthetical problems and physiological disorders such as abnormal cornea exposition and pathological loading of the optic nerve, orbital blood vessels and ocular muscles. It may induce the alteration of visual acuity up to blindness.

Once the endocrinal situation is stabilized, a surgical reduction of the proptosis is usually needed to decompress the orbital content. Two main surgical techniques are actually available. The more recent technique is the "fat removal orbital decompression" (FROD) (Olivari, 1991) and the most commonly used technique seems to be the "bone removal orbital decompression" (BROD) (Ogura & Walsh, 1962, Adenis et al., 2003). It aims at increasing the volume of the orbital cavity with an osteotomy (*i.e.* a resection) of the orbital walls (Stanley et al., 1989, Wilson & Manke, 1991). The soft tissues are partially evacuated through the osteotomy to form a hernia (Fig.1 (b)). The surgeon may manually apply a controlled load on the eyeball. Limited cuts in the outer membrane containing the orbital soft tissues improve the decompression permitting the physiological liquid to flow towards the maxillary and ethmoid sinus regions. The backward displacement of the eyeball to recover the normal position is initiated and it is fully reached after the complete reduction of orbital tissue inflammation. This intervention is technically difficult because of the proximity of ocular muscles and optic nerve, the tightness of the eyelid incision, and the narrowness of the operating field. The surgery must be mini-invasive and specific tools to improve surgical planning could be very helpful.

It is admitted that extracting 1 $cm^3$ of decompressed soft tissue is necessary to induce a backward eyeball translation comprised between 1 mm and 1.5 mm. This clinical result is relative to the FROD technique (Adenis & Robert, 1994, Olivari, 1991). Even if this a-posteriori empiric result is interesting from a clinical point of view, it cannot be used for the surgery planning since it does not take into account the osteotomy size. It is admitted that the relationship between the orbital volume and the orbital pressure is non-linear, but the orbital content is a complicated multiphasic media saturated by fluid. The initial strain energy in the orbit before surgery derives at least from the superimposed effects of the increase of fat volume fraction and fluid pressure. Due to significant coupling effects, predicting the relationship between fluid pressure evolution and proptosis reduction is not straightforward.

To our knowledge, (Luboz et al., 2004) was the first work proposing a biomechanical numerical model to assist in the planning of the proptosis reduction. In this preliminary work, a geometrical model and a 3D poroelastic FE model have been presented and validated in a clinical test. The biomechanical model has been evaluated comparing the proptosis reduction in one patient. In a second step (Luboz et al., 2005), predicted and physiological stiffness of the orbital content before and after surgery were quantified. For both approaches discrepancies were under 4 % which is satisfying for such a multi-parameter clinical problem.

From a clinical point of view, the main goal of the proptosis reduction is to optimize the osteotomy size to obtain a suitable eyeball backward displacement proceeding to a mini-invasive surgery. In this paper we hypothesized that the initial and validated predictive numerical FE model reinforced by sensitivity studies of a patient-related data base could help to establish patient-dependant rules for the pre-operative planning of the proptosis surgery. To proceed, a series of 12 patients has been studied and the relationships between the osteotomy size, the targeted backward translation of the eyeball and the patient-related orbit morphology have been established. The clinical data-base has been built on pre-operative CT scans.

**2. Methods**

*2.1 The reference FE model of the orbital content*

In the preliminary study (Luboz et al., 2004), a poroelastic FE model (Fig.2 (a)) has been implemented to predict the proptosis reduction (Marc/Mentat package MSC Software ©, Santa Ana, CA, USA). It has been validated in one clinical application. The patient-related morphology has been derived from pre-operative and post-operative computed tomography (CT) scans and a three-dimensional reconstruction. Material properties have been updated from the clinical measurements (load and displacement). The reference hexahedron meshing (6948 nodes and 1375 elements) was automatically generated from the manual contour segmentation and a homogeneous poroelastic material has been chosen for the orbital soft tissues (Young modulus: 20 kPa, Poisson ratio: 0.1, permeability: 300 mm$^4$/N.s, porosity: 0.1). The boundary conditions concerned the porous phase and the fluid flow (Biot, 1941) at the orbital bony walls and at the osteotomy location (see Fig.2 (a)). An initial fluid pressure (10 kPa) in the pathologic tissue and the surgeon mechanical action on the eyeball (12 N measured) were applied. The eyeball backward translation was predicted considering the eyeball as a rigid body and the volume of decompressed tissue was computed as the difference between meshes before and after loading.

In this previous study (Luboz et al., 2004), the eyeball backward displacement estimation showed a very satisfying discrepancy of 2 % with the clinical result. In addition, the osteotomy area had a preponderant influence on the backward displacement.

*2.2 Automatic generation of new patient meshing*

The patient database included 12 adults: 7 males and 5 females. All were struck down with proptosis due endocrinal dysfunction. An automatic meshing method based on the mesh-matching algorithm (Couteau *et al.*, 2000) has been used to generate eleven patient meshes from the clinical data base using the initial validated model, previously described, as a reference (see Fig.2 (b)). The patient-related orbital surface was derived from preoperative segmented CT scans data. In order to fit this surface, the reference mesh is deformed using an elastic transformation combined with an optimization procedure. The procedure required only a few minutes to generate the patient-related model and permitted to save the topology of the mesh: number of elements and nodes, zone for controlled boundary conditions. The method was robust despite of rather large variations of shape and orbital volume observed in the data base of 12 patients.

*2.3 Design of the patient-related sensitivity analysis*

The orbital morphology has been characterized by three parameters: the opening radius of the orbit $r$, the depth of the orbit $h$ and the volume of the orbit that was quantified from the 3D mesh. The average volume was 25 cm$^3$ and rather large variations up to ± 26 % have been observed.

As described in (Luboz *et al.*, 2004), the osteotomy size showed a greater influence than its location on the eyeball backward displacement. To investigate this clinical parameter a sensitivity study involving four sizes of osteotomy has been initially designed with the reference FE model and implemented into the 11 other patients thanks to the automatic mesh generation. The standardized sizes, $s$, of the osteotomy (osteotomy area divided by the overall orbit area) were 1.2% ( $s_1$ = 0.8 cm$^2$, standard deviation, SD: 0.1 cm$^2$), 2.8% ( $s_2$ = 1.7 cm$^2$, SD: 0.18 cm$^2$ ), 5.5% ( $s_3$ = 3.4 cm$^2$, SD: 0.31 cm$^2$ ) and 9.6% ( $s_4$ = 5.9 cm$^2$, SD: 0.6 cm$^2$ ). The size of osteotomies has been chosen according to the clinical feasibility in cooperation with the surgical team involved in the study.

Other boundary conditions and loading, described in section 2.1, remained the same for each patient and the control in the FE model data base was straightforward thanks to the saving of the mesh topology and numbering. Finally, 4 FE computations for each of the 12 patients (7 males and 5 females) was achieved to quantify in a first step, the influence the osteotomy size ($s_i$) both on the volume of decompressed tissue ($v_t$) and the eyeball backward translation ($x$). In a second step, the patient-related orbital volume $v_o$ was introduced. Even if the volumes were available from the 3D reconstruction, the simplified cone-like orbit (opening radius $r$, depth $h$ and $v_o = \pi r^2 h/3$) improved the applicability in a clinical setting. A simple radiography or a slice of the CT scan can easily provide $r$ and $h$.

To summarize, there was one group of patients: 7 males and 5 females; the pre-operative CT scan were done for all of these 12 patients, the clinical *rule 1* (equation 1, figure 4a) was derived from 48 FE computations (4 osteotomies for each 12 patients) and the clinical *rule 2* (equation 2, figure 4b) was derived using the same 48 FE computations involving the simplified volume $v_0$.

## 3. Results

Each of the 48 computations (4 osteotomies for 12 patients) required approximately 1 hour using a PC (1.7GHz, 1Go, Marc/Mentat package from MSC Software ©).
The relationships between the volume of decompressed tissue ($v_t$), the eyeball backward translation ($x$) and the patient-related orbital volume ($v_o$) are plotted in Fig.3 (a) and Fig.3 (b). The results were globally linearly distributed and the largest discrepancies appeared for the largest sizes of osteotomies ($s_3$ and $s_4$). The sensitivity corresponding to the coefficient of the straight interpolation line was as follows: 2.6 x$10^{-2}$ ($s_1$), 2.6 x$10^{-2}$ ($s_2$), 3 x$10^{-2}$ ($s_3$) and 7.7 x$10^{-2}$ ($s_4$) for the volume of decompressed tissue and 5.1 x$10^{-2}$ ($s_1$), 6.9 x$10^{-2}$ ($s_2$), 9.2 x$10^{-2}$ ($s_3$) and 10.7 x10$^{-2}$ ($s_4$) for the eyeball migration.. The sensitivities increased with the osteotomy size and all the slopes were positive. For a same osteotomy size, the greater the patient-related orbital volume was, the greater the range of conceivable backward displacement was.

To go further into the analysis of the osteotomy size, Table 1 summarizes the average value and standard deviation of eyeball displacements and volumes of decompressed tissue for the patient database. As shown in Fig.4 (a), the osteotomy size *s* was non-linearly dependant from the eyeball displacement *x*. In equation (1) this was expressed as an exponential regression rule (*rule 1*). The root mean square (RMS) value for each osteotomy surface was 0.06 cm$^2$ for $s_1$, 0.19 cm$^2$ for $s_2$, 0.82 cm$^2$ for $s_3$ and 0.79 cm$^2$ for $s_4$, which corresponded, in comparison to the osteotomy surface, to discrepancies of 7.5% for $s_1$, 11% for $s_2$, 24% for $s_3$ and 13% for $s_4$.

$$s = e^{(0.9x-1.73)} \tag{1}$$

As written in equation (2) (*rule 2*) and plotted in Fig.4 (b), the osteotomy size was expressed as a bi-quadratic form of *x* and $v_o$ using a least square interpolation method. The RMS value for each osteotomy surface was 0.12 cm$^2$ for $s_1$, 0.26 cm$^2$ for $s_2$, 0.33 cm$^2$ for $s_3$ and 0.47 cm$^2$ for $s_4$. In comparison with the osteotomy area, the discrepancies were 15% for $s_1$, 15% for $s_2$, 10% for $s_3$ and 8% for $s_4$. A better 3D surface fitting with obtained with larger osteotomies which explained the decrease of the RMS value.

$$s = -4.69 + 0.68x + 0.26v_o + 0.01x.v_o + 0.11x^2 - 0.005v_o^2 \tag{2}$$

## 4. Discussion & Conclusion

Combine a clinically validated FE model to sensitivity studies in a patient database permitted to obtain two clinical process rules linking the osteotomy size to the targeted eyeball backward displacement. This result was in agreement with our initial hypothesis. The reliability the biomechanical model was partially verified measuring the overall stiffness of the orbital content [Luboz V. et al. 2004]; [Luboz V. et al. 2005].

When results in Table 1 were examined, it showed that the average value of the normalized standard deviation of osteotomy size ($SD_{si}/s_i$) was 11%; the average value of the normalized $SD$ of eyeball backward displacement ($SD_{xi}/x_i$) was 16% and the average value of the normalized $SD$ of decompressed tissue volume ($SD_{vti}/v_{ti}$) was 9%. These overall results permitted to globally quantify the influence of morphology variation associated to per-operative variation of the osteotomy size on the planned exophtalmia reduction.

It appeared that the evaluation of the orbit volume was useful to improve the surgical planning. As shown in Fig.3 (b) the surgeon latitude was increased in-patient with higher orbital volume and the surgical output was altered in patients with lower orbital volume. Even if *rule 1* (equation 1) provided a useful first indication, *rule 2* (equation 2) taking into account the patient-related orbital volume seemed more relevant as it showed half of the RMS value of *rule 1*. Simply describing the orbit as a cone permitted to avoid the segmentation and the 3D reconstruction from CT scans. Moreover, the cone parameters ($r$, $h$) can be easily measured from a single sagittal radiography or a pre-operative CT scan.

It must also be kept in mind that some "optimal" eyeball backward displacement could be physiologically unrealistic since being in conflict with the surgery feasibility (osteotomy size). This potential limitation of the surgical technique would be predicted by the non-linear process rules (*rule 1* and *rule 2*) and the osteotomy size could easily be managed using optical guided medical tools.

The core of the process was a poroelastic FE model validated in a specific clinical and research protocol (Luboz *et al.*, 2004). The sensitivity studies of the current study leant on a 12 patient database from which only pre-operative CT scans were recorded. The fat - muscle ratio and the quality of orbital tissue could have been relevant for the pre-operative planning and the imaging tools to separately reconstruct muscles and optical nerves were available and more complex models could have been done. But the reliability of these models would have been drastically limited by the ignorance of in-vivo material properties (fat, muscle, nerves). So increasing the level of geometric description of the orbital components would not significantly increase the robustness of the predicted results, and surely would limit the applicability in a clinical setting. So the choice was made to model the orbital content as a homogenous biphasic material (fluid/structure) to derive a macroscopic biomechanical behaviour useful for the surgical planning.

The material properties updated and clinically validated in the reference FE model was kept constant for all other patient related models of the database but some clinical observations reported that the orbital soft tissue behavior varies (Gas, 1997). A rheological approach has been initiated in the reference patient to measure the orbital stiffness (Luboz *et al.*, 2005; Payan *et al.*, 2003). The influence of the initial fluid pressure in the orbital content will also be evaluated.

The preliminary post-operative results provided encouraging results and they seemed to confirm our compromise between accuracy and modelling difficulties in the clinical setting (Luboz V. et al. 2004; Luboz V. et al. 2005). But only a statistical pre and post operative study will permit to establish the robustness of the clinical rules presented in the paper. This procedure will require a legal agreement for post operative CT exams which are not acquired in the current clinical routine. This complementary approach will be applied to an enlarged patient database to rank the influence of the pathology level on the surgical output, which would lead to balance the surgical process rules proposed in this article.

*Computer assisted planning and orbital surgery: patient-related prediction of the osteotomy size in proptosis reduction.*
*Vincent Luboz, Dominique Ambard, Pascal Swider, Frank Boutault, Yohan Payan*

**Captions for illustrations**

**Figure 1**. CT scan of a bilateral exophtalmia and image segmentation

**Figure 2**. (a) Reference poroelastic FE model, (b) patient-related model using the mesh-matching algorithm

**Figure 3.** Influence of the orbital volume $v_o$: (a) sensitivity of the volume $v_t$ of decompressed tissue, (b) sensitivity of the eyeball backward translation $x$.

**Figure 4**. Surgical process rules for the osteotomy size: (a) *rule 1* (exponential function), (b) *rule 2* (bi-quadratic function).

*Computer assisted planning and orbital surgery: patient-related prediction of the osteotomy size in proptosis reduction.*
*Vincent Luboz, Dominique Ambard, Pascal Swider, Frank Boutault, Yohan Payan*

**Table 1.** Averaged value of eyeball backward displacement $x$ and volume of decompressed tissue $v_t$ from the patient database.

| $s_i$ (cm²) | $SD_{si}$ (cm²) | $x_i$ (mm) | $SD_{xi}$ (mm) | $v_{ti}$ (cm³) | $SD_{vti}$ (cm³) |
|---|---|---|---|---|---|
| 0.8 | 0.1 | 1.5 | 0.28 | 1.0 | 0.1 |
| 1.7 | 0.18 | 2.5 | 0.39 | 1.7 | 0.17 |
| 3.4 | 0.31 | 3.5 | 0.54 | 2.4 | 0.16 |
| 5.9 | 0.6 | 3.8 | 0.57 | 3.6 | 0.32 |

*Computer assisted planning and orbital surgery: patient-related prediction of the osteotomy size in proptosis reduction.*
*Vincent Luboz, Dominique Ambard, Pascal Swider, Frank Boutault, Yohan Payan*

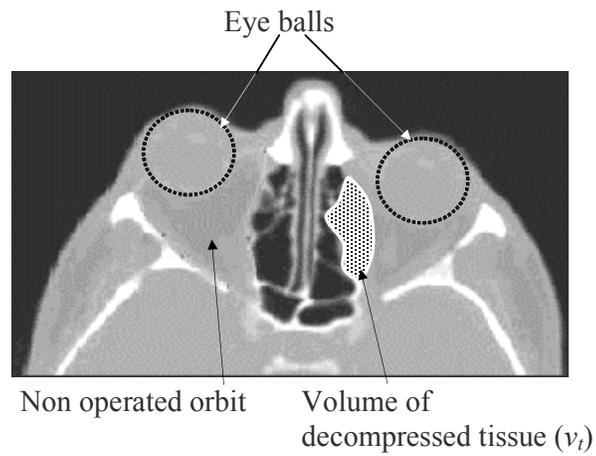

**Figure 1.** CT scan of a bilateral exophtalmia and image segmentation

*Computer assisted planning and orbital surgery: patient-related prediction of the osteotomy size in proptosis reduction.*
*Vincent Luboz, Dominique Ambard, Pascal Swider, Frank Boutault, Yohan Payan*

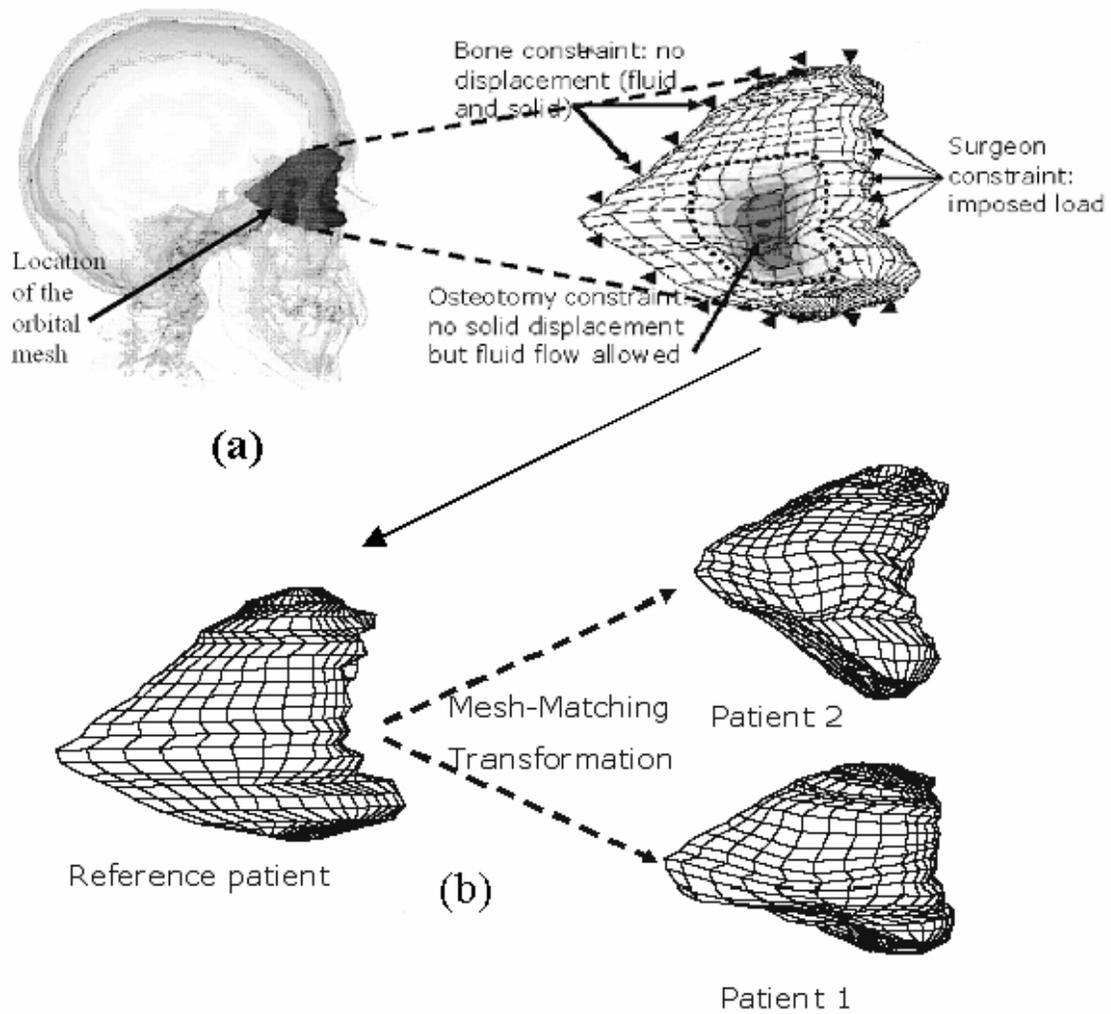

**Figure 2.** (a) Reference poroelastic FE model, (b) patient-related model using the mesh-matching algorithm

*Computer assisted planning and orbital surgery: patient-related prediction of the osteotomy size in proptosis reduction.*
*Vincent Luboz, Dominique Ambard, Pascal Swider, Frank Boutault, Yohan Payan*

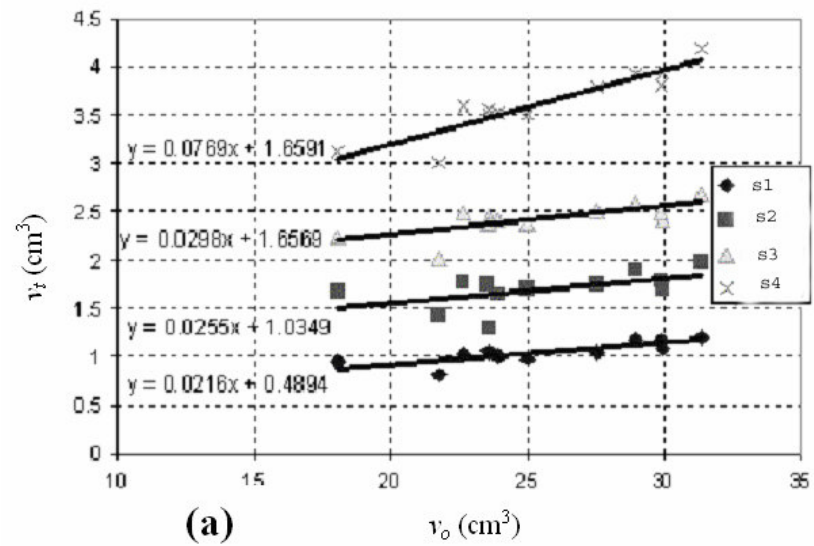

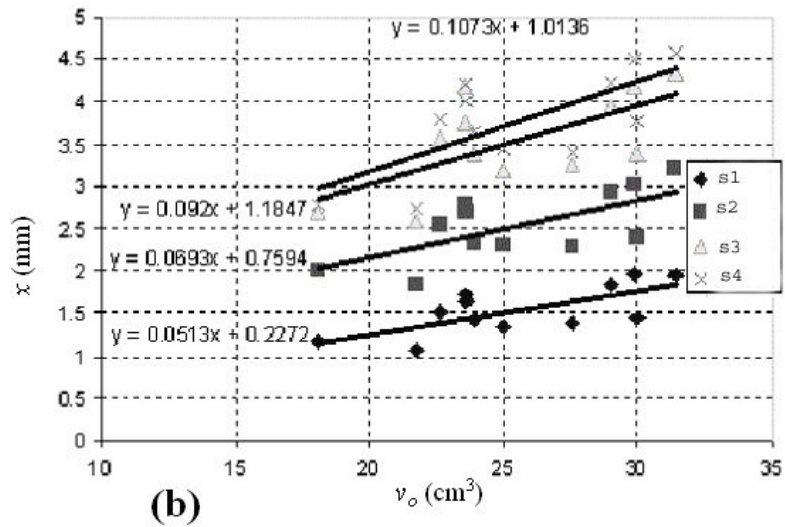

**Figure 3.** Influence of the orbital volume $v_o$: (a) sensitivity of the volume $v_t$ of decompressed tissue, (b) sensitivity of the eyeball backward translation $x$.

*Computer assisted planning and orbital surgery: patient-related prediction of the osteotomy size in proptosis reduction.*
*Vincent Luboz, Dominique Ambard, Pascal Swider, Frank Boutault, Yohan Payan*

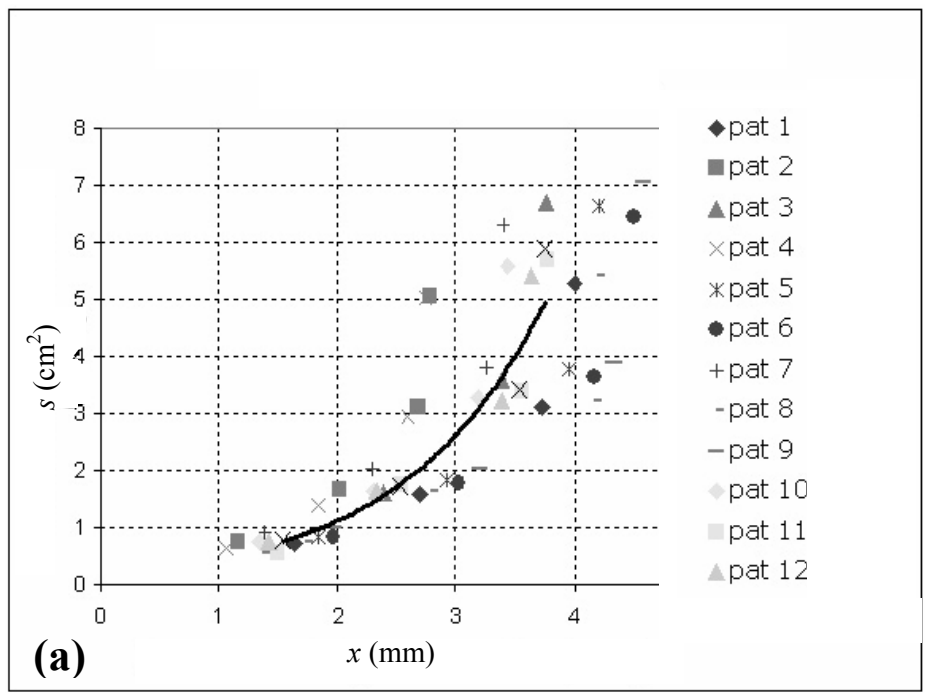

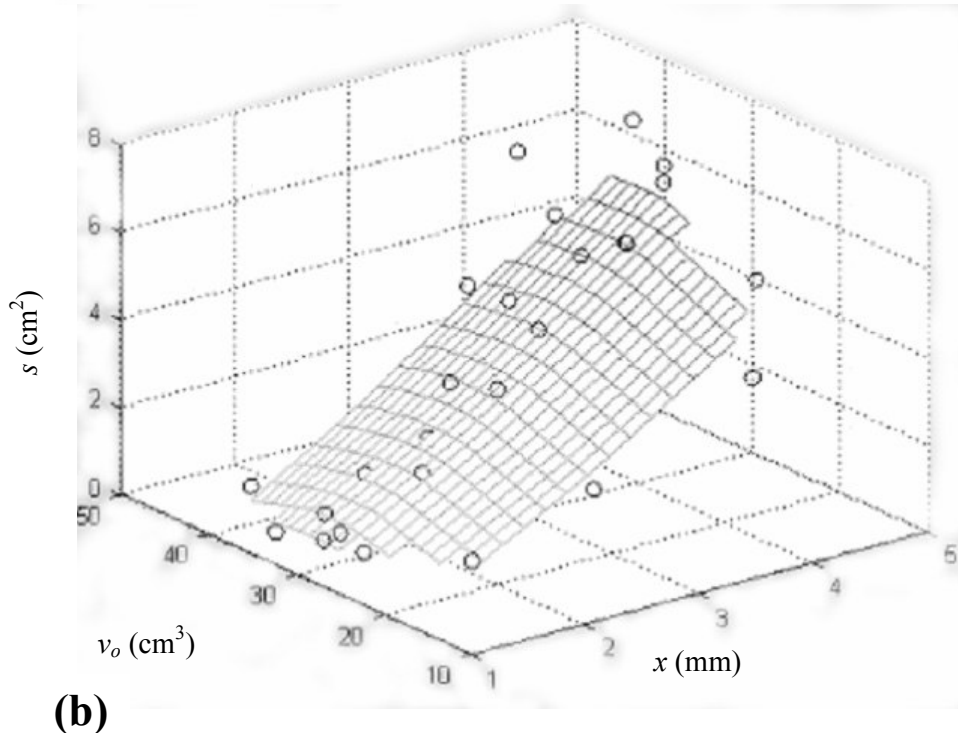

**Figure 4.** Surgical process rules for the osteotomy size *s*: (a) *rule 1* (exponential function), (b) *rule 2* (bi-quadratic function)